\documentclass[journal]{IEEEtran}
\usepackage{cite}
\usepackage{amsmath,amssymb,amsfonts}
\usepackage{tabularx}
\usepackage{blindtext}
\usepackage{graphicx}
\usepackage{textcomp}
\usepackage{wrapfig}
\usepackage{multicol}
\usepackage{multirow}
\usepackage{color,colortbl}
\usepackage[normalem]{ulem}
\usepackage{paralist}
\definecolor{mygray}{rgb}{0.8, 0.8, 0.8}
\newcommand{\hlgray}[1]{{\cellcolor{mygray}#1}}

\def\BibTeX{{\rm B\kern-.05em{\sc i\kern-.025em b}\kern-.08em
    T\kern-.1667em\lower.7ex\hbox{E}\kern-.125emX}}
\definecolor{abstractbg}{rgb}{0.89804,0.94510,0.83137}
\graphicspath{{imm/}}
\usepackage{makecell} 
\usepackage{algorithm}
\usepackage{algpseudocode}
\usepackage{array,booktabs}
\usepackage{siunitx} 
\usepackage{soul}
\usepackage{verbatim}
\usepackage{pifont}

\usepackage[printonlyused]{acronym}

\usepackage{hyperref}
\hypersetup{
     colorlinks = true,
     linkcolor = blue,
     anchorcolor = blue,
     citecolor = blue,
     filecolor = blue,
     urlcolor = blue
}

\begin{document}
\bstctlcite{IEEEexample:BSTcontrol}
\acrodef{AAL}{Ambient Assisted Living}
\acrodef{AmI}{Ambient Intelligence}
\acrodef{GPS}{Global Positioning System}
\acrodef{AP}{Access Point}
\acrodef{RSSI}{Received Signal Strength Indicator}
\acrodef{RP}{Reference Point}
\acrodef{FoG}{Freezing of Gait}
\acrodef{PD}{Parkinson's Disease}
\acrodef{AD}{Alzheimer's disease}
\acrodef{BLE}{Bluetooth Low Energy}
\acrodef{IoT}{Internet of Things}
\acrodef{ADL}{Activities of Daily Living}
\acrodef{ICT}{Information and Communication Technology}
\acrodef{RF}{Radio Frequency}
\acrodef{WLAN}{Wireless Local Area Network}
\acrodef{RFID}{Radio Frequency IDentification}
\acrodef{UWB}{Ultra Wide Band}
\acrodef{AOA}{Angle of Arrival}
\acrodef{DOA}{Direction of Arrival}
\acrodef{TDOA}{Time difference of Arrival}
\acrodef{TOA}{Time of Arrival}
\acrodef{MV}{Magnetic Field Vector}
\acrodef{GNSS}{Global Navigation Satellite System}
\acrodef{PS}{Positioning System }
\acrodef{VS}{Voting System}
\acrodef{ML}{Machine Learning}
\acrodef{k-NN}{k-Nearest Neighbour}
\acrodef{SVM}{Support Vector Machine}
\acrodef{VCS}{Virtual Coaching System}
\acrodef{HAR}{Human Activity Recognition}
\acrodef{AR}{Activity Recognition}
\acrodef{HVAC}{Heating, Ventilating, and Air-Conditioning}
\acrodef{ODL}{Objects of Daily Life}
\acrodef{SBC}{Single Board Computer}
\acrodef{PIR}{Passive InfraRed}
\acrodef{RMSE}{Root Mean Square Error}
\acrodef{MAE}{Mean Absolute Error}
\acrodef{MSE}{Mean Squared Error}
\acrodef{SVM}{Support Vector Machine}
\acrodef{kNN}{k-Nearest Neighbor}
\acrodef{WkNN}{Weighted kNN}
\acrodef{LDA}{Linear Discriminant Analysis}
\acrodef{QLDA}{Quadratic LDA}


\title{Estimating indoor occupancy through low-cost BLE devices}
\author{Florenc Demrozi, \IEEEmembership{Member, IEEE}, Cristian Turetta, \IEEEmembership{Member, IEEE}, Fabio Chiarani, Philipp H. Kindt, and Graziano Pravadelli, \IEEEmembership{Senior Member, IEEE}
\thanks{
This research work has been partially supported by the
project Dipartimenti di Eccellenza 2018-2022 funded by
the Italian Ministry of Education, Universities and Research
(MIUR).\newline
(Corresponding author: Florenc Demrozi.) }
\thanks{F. Demrozi, C. Turetta, F. Chiarani, and G. Pravadelli are with the Computer Science Department, University of Verona, Italy (e-mail: name.surname@univr.it).
P. Kindt is with the Faculty of Computer Science, Chemnitz University of Technology, Germany (e-mail: philipp.kindt@informatik.tu-chemnitz.de).
}}


\maketitle
\begin{abstract}
Detecting the presence of persons and estimating their quantity in an indoor environment has grown in importance recently. 
For example, the information if a room is unoccupied can be used for automatically switching off the light, air conditioning, and ventilation, thereby saving significant amounts of energy in public buildings. 
Most existing solutions rely on dedicated hardware installations, which involve presence sensors, video cameras, and carbon dioxide sensors. 
Unfortunately, such approaches are costly, are subject to privacy concerns, have high computational requirements, and lack ubiquitousness. 
The work presented in this article addresses these limitations by proposing a low-cost occupancy detection system.
Our approach builds upon detecting variations in Bluetooth Low Energy (BLE) signals related to the presence of humans. 
The effectiveness of this approach is evaluated by performing comprehensive tests on five different datasets.
We apply several pattern recognition models and compare our methodology with systems building upon IEEE 802.11 (WiFi).
On average, in multifarious environments, we can correctly classify the occupancy with an accuracy of 97.97\%. 
When estimating the number of people in a room, on average, the estimated number of subjects differs from the actual one by 0.32 persons. 
We conclude that our system's performance is comparable to that of existing ones based on WiFi, while significantly reducing cost and installation effort. 
Hence, our approach makes occupancy detection practical for real-world deployments.
\end{abstract}
\begin{IEEEkeywords}
\ac{BLE}, Occupancy detection , Occupancy counting, Pattern recognition. 
\end{IEEEkeywords}
\section{Introduction}\label{sec:intro}
The deployment of smart buildings has gained momentum in recent years, thanks to the wide availability of low-cost and accurate sensing and actuating devices, which are controlled by advanced artificial intelligence-based systems. This has promoted the development of several smart applications in many relevant scenarios, like, for example, health care~\cite{acampora2013survey,tan2018exploiting}, assistance of  elderly and people with special needs~\cite{demrozi2019towards,rashidi2013survey,demrozi2019indoor}, human activity recognition~\cite{demrozi2020human,chen2018building}, heating, ventilation, and air-conditioning systems~\cite{ehlers2006system}, recognition of the environmental status~\cite{pires2018recognition}, and smart home solutions in general~\cite{cook2003mavhome}.

One of the critical problems in these scenarios is detecting whether a room is occupied. If this information is available in real-time, relevant decisions can be taken quickly and automatically. 
For example, air-conditioning/heating and lighting can be switched off once a room is empty. 
If the number of people in a room can be estimated, systems such as ventilation can be controlled adaptively, thereby improving the air quality and further reducing the energy consumption. The occupancy information is also relevant in case of an emergency, when rescue actions should be directed towards rooms in which the presence of people has been detected.
Hence, automatically detecting the presence of people and estimating their quantity has been studied frequently in recent times~\cite{yang2018comparison,golestan2018data,zou2018device,GGK20}. \\


\par\noindent\textbf{Related work: }Technologies for occupancy detection and estimation of the number of people in a room can be broadly categorized into a) room installations, e.g.,~\cite{zou2018device}, and b) body-worn devices, e.g.,~\cite{yang2018comparison}.
Body-worn solutions rely on sensors that are worn on the body of each subject in a room.  These devices emit wireless signals, which are analyzed by receivers installed in the environment.
However, the assumption that every person always wears such a device when entering a specific room is unrealistic in practical scenarios. 
For this reason, recent research has focused on techniques that rely only on devices deployed within the room infrastructure~\cite{trivedi2019occupancy,chen2018building}, without requiring any body-worn devices. 
Techniques that solely rely on room installations exploit analyzing properties such as, e.g., the CO$_2$ level, audio signals, video images, or radio signal propagation patterns.
Other feasible methods rely on indirect measures, such as the air quality, which can be assessed using electromagnetic fields~\cite{hasenfratz2013spatially}. 
In addition, a large number of published approaches rely on inferring occupancy information from the Channel State Information (CSI) of WiFi networks.
Occupancy detection using wireless signals is a challenging problem, especially because of the multitude of activities and positions that people can occupy within an environment, leading to a large variety of different signal patterns.
For example, Yang et al.~\cite{yang2018device} presented a real-time, device-free, and privacy-preserving WiFi-enabled IoT platform for occupancy sensing. 
This approach makes use of CSI information and can achieve a detection accuracy of 96.8\%. 
Depatla et al.~\cite{depatla2015occupancy} propose a methodology to identify two distinctive patterns in the CSI data related to people in a room: blocking the Line Of Sight (LOS) and scattering.
Based on this, it is possible to estimate the total number of occupants, thereby achieving an accuracy of 96\%.
Similarly, Zou et al.~\cite{zou2018device} presented an occupancy detection methodology based on CSI data. It measures the similarity between adjacent CSI time series and reaches an accuracy of 99.1\%. 
Chen et al.~\cite{chen2013non} proposed an occupancy detection system, which relies on analyzing the changes in the statistical metrics of the power consumed by the building (i.e., the electricity for ventilation and lightning).
Similarly, Akbar et al.~\cite{akbar2015contextual} presented an occupancy detection system based on the electric power consumption. Here, machine learning algorithms, such as kNN and SVM were applied. This approach claims an average accuracy of 94\%.
Furthermore, BLE-based systems have been widely studied. 
Mateos et al.~\cite{mateos2020tool} presented a body-worn methodology that uses BLE beacons with a broadcast frequency of $\approx$ 10 Hz and smartphones to determine the level of occupancy in indoor and outdoor spaces, thereby achieving an average accuracy of 95\%.
Chen et al.~\cite{chen2019collecting} presented a stochastic methodology that simultaneously uses BLE and WiFi data. It uses a frequency of 1 Hz and utilizes connection/traffic information to estimate the number of users and the environment status. \\

\par\noindent\textbf{Limitations of existing approaches:} Most of the previously known approaches require special hardware with high computational power. 
In addition, dedicated devices have to be installed into the rooms under surveillance. 
This also holds true for WiFi-based setups. In particular, commercially available WiFi devices, such as routers and laptops, typically do not provide access to CSI data. Only very few WiFi Systems on Chips (SoCs) natively provide access to the CSI data, and all of them are deprecated. 
For this reason, most recent approaches build upon Broadcom's BCM43 series, for which a firmware patching framework~\cite{nexmon:project, gringoli:19} unlocks access to the CSI data. 
Hence, special WiFi receivers need to be installed into a building only for the purpose of gathering CSI data. 
Since WiFi radios are typically power-hungry, they need access to the electricity grid. In addition, the patched firmware prevents such devices from actively communicating over WiFi networks, since they can only act as observers that extract WiFi signals. 
Hence, such receivers also require access to a wired network for relaying the gathered CSI data to a server. In summary, occupancy detection systems based on CSI incur a significant installation effort and hence cost. 
Furthermore, they need to be planned and installed in advance and cannot be flexibly used on-the-fly when the necessity of occupancy detection arises at short notice.
Finally, the main limitation of existing BLE-based methodologies is that they are mostly integrated into body-worn systems interacting with an existing WiFi/BLE infrastructure. In addition, their sampling frequency is usually lower than $\SI{10}{ Hz}$, which negatively impacts their accuracy.\\

\par\noindent\textbf{Contributions of this paper:}
To overcome these limitations, this paper presents an easily accessible occupancy detection and people counting platform, based on a mobile Android application and BLE devices (e.g., using the low-cost nRF52832 SoC).
The proposed system, which relies on a few \ac{BLE} devices, can be flexibly deployed in a room, e.g., by gluing them onto the walls. 
Such devices can be battery-powered and might operate for multiple days to weeks before needing to be recharged. 
Hence, they do not need any access to the electricity grid and neither need to be connected to the wired network. 
An Android smartphone, a Single Board Computer (SBC) or a PC, which can send the occupancy results to a server via WiFi, acts as a BLE receiver. 
Our detection system analyzes the received BLE signals using pattern recognition techniques. 
It is driven by the insight that occupancy causes variations in the radio signal propagation patterns, which can be observed in the Received Signal Strength Indicator (RSSI).
The main characteristics of the proposed approach are the following ones.
\begin{compactitem}
    \item Low-cost: Each node costs only a few USD (e.g., the nRF52832 SoC costs $\approx$ 2 USD), which is only a fraction of the cost of a WiFi AP (i.e., $\approx$ 100 USD).
    The receiver can be a standard Android smartphone, which are available starting from below 100 USD;
    \item Non-invasive: Users do not have to carry any devices on their body and cameras are not used;
    \item Compatible: Our approach works properly with any BLE sender/receiver that provides the ability of measuring the RSSI with a frequency of at least 45 Hz;
    \item Ubiquitous and flexible: Being a mobile, ''pocket-size'' system, our approach is suitable for environments without  existing infrastructure;
    \item Accurate: The performance of the proposed approach is comparable to that of existing CSI-based systems.
\end{compactitem}  
~\\
Overall, the main contributions of this work are as follows.
\begin{compactitem}
\item We propose a BLE-based occupancy detection system, which comes with considerably lower cost and installation effort than existing approaches;
\item We evaluate its performance by using multiple real-world measurements and by applying a workflow exploiting several pattern recognition algorithms (i.e., regression and classification algorithms), fed by both feature representations and raw measurement data;
\item We experimentally compare its performance to state-of-the art CSI-based systems. To make both approaches comparable, we implemented such a CSI-based system in the same environments as our proposed, BLE-based one. Our results suggest that the much simpler RSSI signal of \ac{BLE} is sufficient for both accurately detecting the occupancy of a room, as well as for estimating the number of people inside of it.
\end{compactitem}

\par\noindent\textbf{Our own previous work: }
This paper is an extension of our conference paper~\cite{demrozi2021date}. Among other changes, we have added an evaluation that compares our BLE-based approach to a method that exploits state-of-the-art CSI information.\\

\par\noindent\textbf{Organization of the paper: }The rest of the paper is organized as follows. Section \ref{sec:back} provides the necessary background. Section~\ref{sec:over} details the proposed methodology. Section~\ref{sec:res} discusses experimental results. Finally, concluding remarks are reported in Section~\ref{sec:conc}.

\section{Preliminaries}\label{sec:back}
This section introduces the necessary background, on which our proposed approach is built upon.
\subsection{Received signal strength indicator (RSSI)}
In radio communication technologies, the \ac{RSSI} indicates the received signal power measured by the receiver device.  
In \ac{BLE}, the \ac{RSSI} is an integer value that indicates the received power in dBm. The \ac{RSSI} is often exploited for fingerprinting~\cite{khalajmehrabadi2017modern} in localization applications, where a certain signature of \ac{RSSI} values is used to identify a known location. 
In this paper, we exploit different RSSI propagation patterns to identify whether a room is occupied and estimate the number of people inside of it.
The analysis of the gathered data through dedicated pattern recognition techniques provides the capability to identify patterns that correspond to the number of people in the environment and, in some cases, to the activities that these people are carrying out (e.g., walking, laying down, or sitting)
~\cite{yang2018device}.


\subsection{Channel state information}
Following the 802.11ac standard, WiFi networks use Orthogonal Frequency Division Multiplexing (OFDM) as a digital transmission method.
Here, every channel utilizes a bandwidth of 20, 40, 80, or 160 MHz~\cite{rhodeSchwarz}.
Each channel is subdivided into 64 (for 20 MHz channels) to 512 (for 160 MHz channels) subcarriers. Each subcarrier uses distinct frequencies within the channel, and data is transmitted simultaneously on all of these subcarriers in parallel.
Let $X$ be the signal emitted on a specific subcarrier, and $Y$ the corresponding signal that has arrived at the receiver.
Then, typically the following relation is assumed~\cite{wang:19}.
\begin{equation}
    Y = H \cdot X + N
\end{equation}
Here, $N$ represents noise, whereas $H$ is the \textit{channel state information (CSI)}.
When a signal is transmitted over the wireless link, it is attenuated and/or undergoes a phase change. Both effects are quantified by the CSI $H$. Hence, $H$ for a single WiFi frame is a vector that contains a complex number for every subcarrier, describing how the amplitude and phase of the signal have changed. The presence of human bodies and their movements have a major impact on $H$. Therefore, it is possible to infer information on room occupancy from CSI signals.

In our experiments, we use a bcm43455c0 WiFi radio with a modified firmware based on the Nexmon firmware patching framework~\cite{nexmon:project, gringoli:19}. 
We use $\SI{20}{ MHz}$ channels with 64 subcarriers. Hence, for every received WiFi frame, we obtain 64 amplitude and phase values, of which 56 are related to actual data transmissions~\cite{rhodeSchwarz}. 
In this work, we use the amplitude information from multiple different access points in a room to infer whether or not the room is occupied. 
Towards this, we use the classifiers described in Section~\ref{sec:ocounting} and compare the results obtained when using CSI with those obtained from a \ac{BLE}- and \ac{RSSI}-based system.

\subsection{Regression and classification algorithms} 



Regression and classification are artificial intelligence-based techniques, which apply a supervised learning algorithm on labeled training data. After having learned from the labeled data, an unknown input can be classified, or a property that has some dependency on the input data can be predicted~\cite{bishop1995neural}. 


Classification aims to decide which choice among a set of classes explains best a certain, previously unknown input. It is the most common operation in machine learning and depends strongly on the data representation (i.e., raw data, manually or automatically extracted features).
The quality of this classification is commonly assessed by the \textit{precision}~$P$, \textit{specificity}~$S$, \textit{recall}~$R$, also called \textit{sensitivity}, and overall \textit{accuracy}~$A$. These metrics are defined as follows~\cite{powers2011evaluation}:\\
\begin{equation}
\label{eq:qualityMetrics}
{
\renewcommand{\arraystretch}{1.4}
\begin{array}{cc}
P = \frac{ tp}{ tp +  fp} & S = \frac{ tn}{ fp +  tn}\\
R = \frac{ tp}{ tp +  fn} & A  = \frac{ tp +  tn}{ p +  n}
\end{array}
}
\end{equation}
Here, $tp$ represents the number of true positives, $fp$ the number of false positives, $n$ the total number of negatives, and $p$ the total number of positives.
Precision and recall quantify how the classifier can avoid false positives and correctly classify all of the samples that belong to a specific class, while specificity quantifies the ability to classify true negatives correctly. Finally, accuracy is the number of correctly classified data samples out of all samples.

While the goal of classification is determining to which class a certain observation belongs, regression attempts to predict a value based on an observation of the input data. A set of independent variables that form the input are called ``predictors'' or ``features''~\cite{bishop1995neural}.
The accuracy of the regression model is measured in terms of \ac{RMSE} and \ac{MAE}, calculated as given by the following equation~\cite{powers2011evaluation}:\\
\begin{equation}
\label{eq:RMSE_MAE}
\begin{array}{l}
    RMSE = \sqrt{ \frac{1}{n}\sum_{i=1}^{n} (\left | y_{i} - x_{i} \right |)^2}\\
    MAE = (\frac{1}{n})\sum_{i=1}^{n}\left | y_{i} - x_{i} \right | \\
    \end{array}
\end{equation}

Here, $x_i$ identifies the actual outcome of the input data, $y_i$ the estimated outcome, and $n$ the number of samples under consideration.
\begin{table*}[!ht]
\centering
\begin{tabular}{c|c|c|c|c|c|c||c|c}
\hline
\hline
Timestamp&$Mac_{Addr_1}$&$Mac_{Addr_2}$&$Mac_{Addr_3}$&$Mac_{Addr_4}$&$\dots$&$Mac_{Addr_n}$&Occupancy&Nr. Occupants \\
\hline
26/08/2020 09:56:45.005 &-51 &-65 &-80 &-100 &\dots &-35 & true & 2 \\
26/08/2020 09:56:45.010 &-41 &-55 &-70 & -90 &\dots &-45 & true & 4 \\
\dots &\dots &\dots &\dots &\dots &\dots &\dots&\dots&\dots\\
26/08/2020 10:00:00.000 &-37 &-49 &-65 & -70 &\dots &-35 & false& 0 \\
\hline
\hline
Distance (cm)&25 &500 &100  &300 &\dots &600\\
\cline{1-7}
\end{tabular}
\vspace{-0.2cm}
\caption{Dataset of RSSI measurements}\label{tab:coll_data}
\vspace{-0.5cm}
\end{table*}
\section{Methodology}\label{sec:over}
\begin{figure}[!t]
\centering
\includegraphics[width=\columnwidth, page={2}]{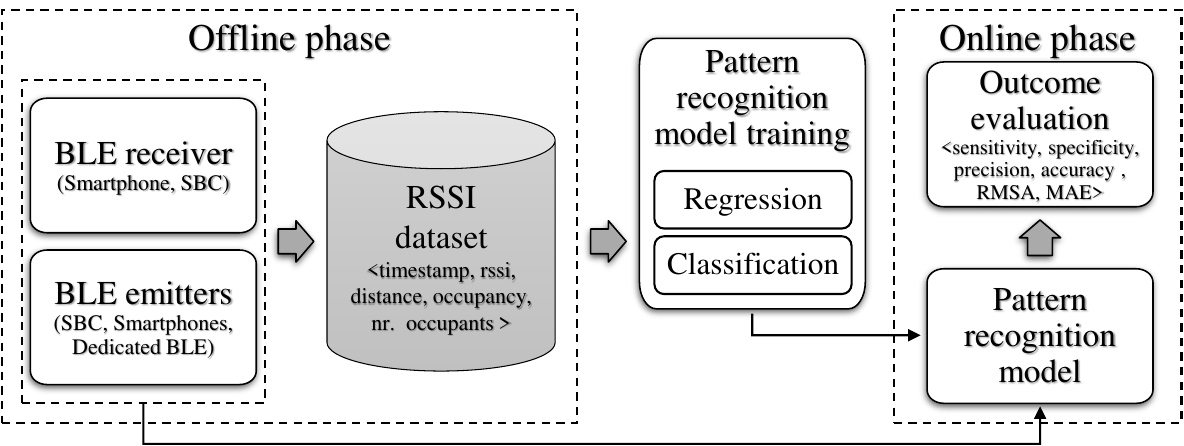}
\caption{Schematic view of the proposed methodology.}\label{fig:detailedview}
\end{figure}

\color{black}
The methodology we propose for occupancy detection and people counting consists of the following three phases (cf. Figure~\ref{fig:detailedview}).
\begin{enumerate}
\item Offline RSSI data collection for the training of the pattern recognition models. 
The RSSI is related to the communication between multiple BLE senders (e.g., smartphones, BLE beacons or SBCs) and a single BLE receiver (e.g., smartphone or SBC);
\item Training of multiple pattern recognition models by using the data obtained from the previous phase;
\item Online evaluation of the pattern recognition models for predicting the occupancy in real-time.
\end{enumerate}
The methodology makes use of a communication architecture exploiting a central receiver (typically a smartphone or an SBC) that collects RSSI data from several BLE devices located in the environment. 
The received data are then forwarded to a database on a server. 
The goal is to study the changes in the radio signal propagation patterns generated between the senders and the receiver to estimate the status of the environment (i.e., environment occupancy, and number of people).
For this purpose, we identify the most appropriate recognition/regression model among multiple choices. 

\subsection{Experimental setup}\label{sec:arch}
Initially, the BLE devices establish a synchronized connection to the receiver. 
The receiver periodically forwards data to a server for offline data processing and for the training of the pattern recognition model. 
Though the approach will work with a large variety of BLE devices, in our experiments, we used \href{https://www.nordicsemi.com/Software-and-tools/Prototyping-platforms/Nordic-Thingy-52}{Nordic Thingy™ 52} nodes as transmitters, which are based on the nRF52832 SoC. 
In our experiments, we evaluated different sampling frequencies of the accelerometer of the Nordic Thingy board in the range between 5 Hz and 200 Hz and transmitted these data via BLE. The system adjusts the data transmission rate (i.e., the connection interval in BLE) accordingly, and hence the choice of the sampling frequency of the acceleration sensor also determines the connection interval.
Our approach supports arbitrary transmitter positions in the environment, with the only restriction that the location cannot vary between the offline training and the online prediction phase. 
We use an Android smartphone as the receiving device. 
Hence, our methodology supports simple ad-hoc installations without requiring any fixed infrastructure, since all devices are battery-powered and do not need access to a wired network.
Considering the requirements of future deployments, Android has been chosen due to its compatibility with several pattern recognition libraries (e.g., Keras, Tensorflow, or Weka). 
Though we have carried out the classification and regression on a computer in our experiments, the resulting system supports deployments in which only the training phase is executed on a server, whereas the detection can be done directly on the smartphone. This eliminates the need for the smartphone to maintain an internet connection.

\subsection{Offline data collection}
During the offline data collection phase, the receiver gathers the data collected by the BLE sensors distributed in the environment at a fixed sampling frequency. Each received packet is timestamped and associated with its corresponding RSSI measurement. 
To reduce the energy consumption for communication, we accumulate the samples received by phone and periodically transmit them to the server in a batch.
Finally, we manually assign the number of persons present in the environment at each timestamp to this data. We also manually enter the distance between each emitter and the receiver.

The resulting labeled dataset for training the classifiers, as stored on the server, is exemplified in Table~\ref{tab:coll_data}.
Each column $MAC_{Addr_{i}}$, $i = 1,2,..,n$, refers to the RSSI measurements for the emitter with MAC address $i$. 
The column \textit{occupancy} is \textit{false} when the number of people in the environment is 0, and \textit{true} otherwise. 
The column \textit{Nr. Occupants} contains the number of people inside the observed environment at the corresponding timestamp. The number of people inside the environment is provided by an external human observer who updates the occupants' number every time they leave or enter the environment. These labels are used for supervised learning.
The \textit{Distance} row contains the distance between emitter $i$ and the receiver.
RSSI measurements represent the input (i.e., \textit{predictors}) for the pattern recognition models used during the online evaluation. 
Similarly, the occupancy state and the number of people in the environment are the target outcome (i.e., the \textit{outcome variables}) that we attempt to predict. 
In other words, these are the labels for supervised learning.
Finally, to mitigate the effects of a potential ``overfitting'' of the model, we eliminate duplicates present in the dataset, thereby maintaining only their first occurrence. 
\begin{figure*}[!ht]
\centering
\includegraphics[width=1.9\columnwidth, page={3}]{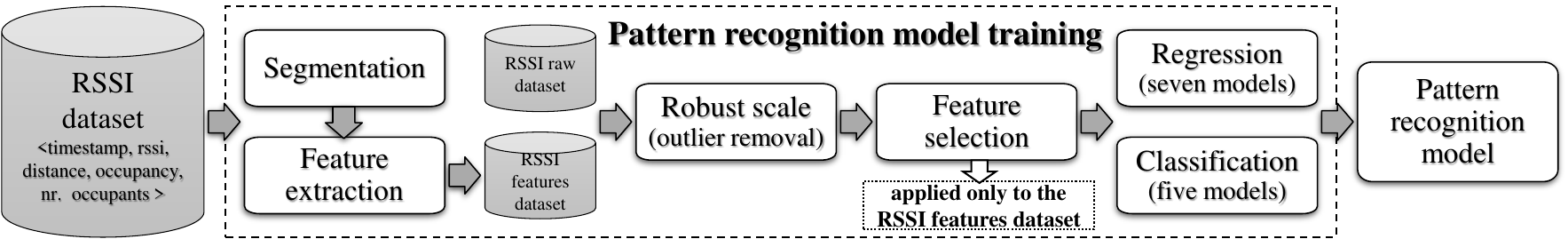}
\vspace{-0.2cm}
\caption{Pattern recognition model training workflow.}\label{fig:training}
\vspace{-0.4cm}
\end{figure*}
\subsection{Pattern recognition model training}\label{sec:ocounting}
\color{black}
In this step, the collected data are used to train a pattern recognition model for the following scenarios:
\begin{enumerate}
\item Recognition of the occupancy status of the environment;
\item Estimation of the number of people inside the environment.
\end{enumerate}
The overall workflow for this phase is shown in Figure~\ref{fig:training}.
The resulting model is capable of predicting the occupancy status and people count online, based on RSSI data from live measurements.
Initially, the collected RSSI data are segmented into time-windows of $\SI{1}{s}$.
Subsequently, from each segment, we extract a comprehensive set of features, as given by Table~\ref{tab:tf_features}, using the software library presented in ~\cite{barandas2020tsfel}. 

We have chosen time windows of 1 second because we assume that the occupancy status cannot change at shorter time scales. For each such window, we compute 159 features per BLE transmitter.
At the end of the segmentation and feature extraction phase, we have obtained two datasets: one that consists of raw RSSI data, which we call the \textit{raw dataset} in the following, and RSSI features data obtained as described above, which we call the \textit{features dataset}. 
Both of them are used in the analysis reported below. Before this analysis, we ``clean''/``normalize'' the data by removing outliers. We thereby use the  \textit{robust scalar}  method, which works as given by the following equation: 
\begin{equation}
\label{eq:Dnorm}
D_{nor} =\left\{x_{nor}: \forall x\in D,~x_{nor}=\frac{x-Q_2^D(x)}{Q_3^D(x) - Q_1^D(x)}\right\}
\end{equation}
Here, $Q_1^D$, $Q_2^D$, $Q_3^D$ are the first, second (aka median), and third quartiles of the dataset $D$, $x$ is a sample of $D$, and $D_{nor}$ is the dataset $D$ after applying the robust scale outlier removal technique. \\

\noindent\textbf{Feature selection:}
Not all extracted features contribute positively to the detection accuracy. 
A higher number of features does not always imply a greater accuracy of the pattern recognition model; similarly, a smaller number of features does not always lead to a reduction of the accuracy. 
Therefore, we identified those features that actually contribute most to the quality of the classification.
For this purpose, we applied feature selection techniques known from the literature~\cite{chandrashekar2014survey}.
The main benefits obtained from the removal of unnecessary/misleading features (e.g., features with very low variance, duplicates of existing features, or high noise) are: i) reduction of overfitting; ii) reduction of noise-related errors; iii) improvement of the accuracy; iv) reduction of the training time~\cite{bishop2006pattern}.
The proposed methodology makes use of a tree-based feature selection technique, which is applied only to the RSSI features dataset.
Tree-based estimators, by definition, internally create an ordering of the features representing the training dataset, which makes them very suitable to be used by feature selection methods. More details on tree-based feature selection can be found in~\cite{chandrashekar2014survey}.
After feature extraction, the training phase takes place. It creates the models for occupancy detection and occupancy counting. More information on the training phase is given in Section~\ref{sec:traval}.
\begin{table*}[!htb]
\centering
\resizebox{\textwidth}{!}{\begin{tabular}{p{8cm} || p{8cm}}
\hline
\hline
Time Domain       & 
Frequency Domain  \\
\hline
\hline
maximum, minimum, mean, standard deviation, root mean square, range, median, skewness, kurtosis, time-weighted variance, interquartile range, empirical cumulative density function,  percentiles (10, 25, 75, and 90), sum of values above or below percentile (10, 25, 75, and 90), 
mean amplitude deviation,
mean power deviation, 
 autoregression.
& 
fast fourier transform (FFT) coefficients, discrete fourier transform (DFT), discrete wavelet transform (DWT), first dominant frequency, ratio between the power at the dominant frequency and the total power, ratio between the power at frequencies higher than 3.5 Hz and the total power, 
wavelet entropy values, 
energy band. \\
\hline
\hline
\end{tabular}}
\caption{Most important time and frequency domain features used in our analysis.}\label{tab:tf_features}
\end{table*}

\subsubsection{Occupancy detection}
Occupancy detection represents a binary classification problem, where \textit{false} means the environment is empty, and \textit{true} that there is at least one person in the environment. 
In particular, five different classification models (i.e., \acf{kNN}, \acf{WkNN}, \acf{LDA}, \acf{QLDA}, and \acf{SVM}) are used. 

Here, the goal is to estimate the occupancy status over a time window of 1 second by using only the RSSI features dataset, since the used classifiers cannot handle raw time series data if they have not been segmented by the segmentation phase shown in Figure~\ref{fig:training}.
The results of this classification problem are evaluated by using the quality metrics we have introduced in Equation~\ref{eq:qualityMetrics}. 

\subsubsection{Occupancy counting}
Unlike occupancy detection, which is a binary classification problem, the occupancy counting scenario is an estimation problem.
Here, we aim to identify the number of people being present in the environment. 
Multivariate regression analysis (i.e., Gradient Boosting, Random Forest, Linear, Ridge, RANSAC, Bayesian, and TheilSen) on both the raw and features dataset predict one variable (viz., the Nr. of occupants), based on  multiple input variables (raw data samples and features)~\cite{bishop1995neural}.
We study two different cases.
In the first case, we only consider raw data. Here, the model returns an estimation of the number of occupants every 5 ms (when a sampling rate of 200 Hz is used).
In the second case, we only consider features data, obtaining an estimation of the number of occupants every second (i.e., once per window length used for data segmentation).  
The results of this prediction are evaluated by using the quality metrics we have introduced in Equation~\ref{eq:RMSE_MAE}.

\subsection{Training and validation}\label{sec:traval}
\begin{figure}[!h]
\centering
\includegraphics[width=0.6\columnwidth, page={4}]{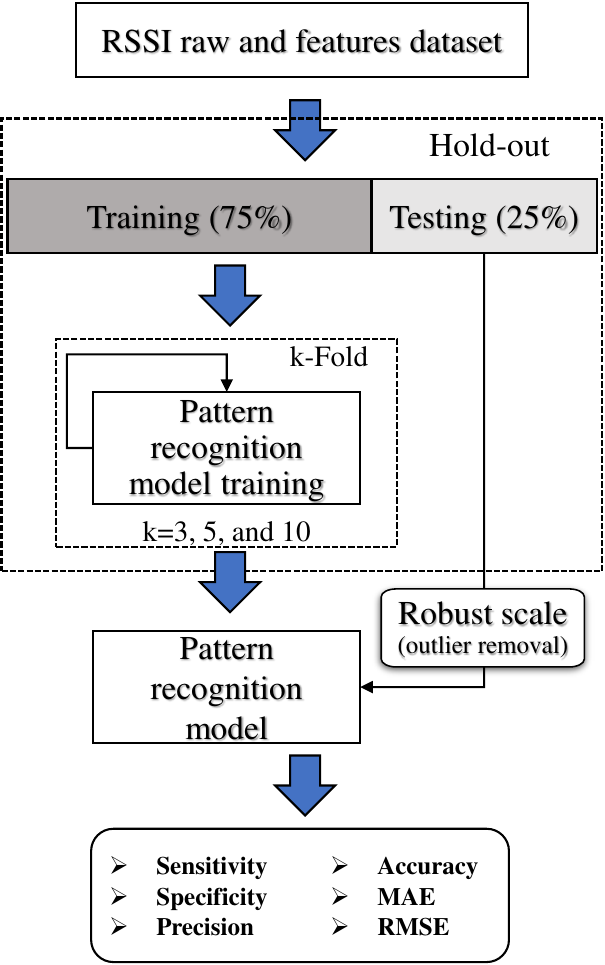}
\caption{Training and validation approach.}\label{fig:test_val}
\end{figure}

To evaluate/validate the quality of our pattern recognition models, the RSSI raw and features datasets are partitioned into \textit{training} and \textit{testing} datasets. 
In particular, for each considered scenario, the raw RSSI and features datasets are initially split (according to a hold-out procedure~\cite{bishop2006pattern}) into a training dataset (75\% of all measurements) and a testing dataset (25\% of all measurements).
On the training dataset, we run the pattern recognition model training phase, as described in Section~\ref{sec:ocounting}. 
We use a $k$-fold cross validation procedure with $k = 3, 5, 10$~\cite{bishop2006pattern}, as shown in Figure \ref{fig:test_val}.
We want to point out that we do not perform the usual model training approach, but we apply the \textit{grid-search} training approach that exhaustively generates optimal pattern recognition model\footnote{with regard to the given grid of parameters} candidates from a given grid of parameters.
For example, the SVM model learns from the training dataset by using different configuration parameters (e.g., kernel function [linear, polynomial, sigmoid, radial~basis~function], penalty [$l_i,~l_2$], or loss function [hinge, squared hinge]). It is tested on the testing dataset for each configuration, returning the configuration that achieved the best results in the training phase.
Finally, we use the testing dataset to further examinate the best model. 
In the end, the procedure returns the above-mentioned evaluation metrics, viz., RMSE, and MAE for regression, and sensitivity, specificity, precision, and accuracy for classification.
\begin{table*}[!ht]
\centering
\resizebox{1\textwidth}{!}{
\begin{tabular}{c|c|c|c|c|c|c|c|c|c}
\hline
Dataset & \# of & Use case & Environment size& Min \#    & Max \#    & Min  & Max &\# of&Size\\
ID& transmitters & scenario &l(m) x w(m) x h(m) & of people & of people & distance (cm)&distance (cm)&samples&(Mb)\\
\hline
\hline
1&6(BLE)&baseline &8.8x8.6x3.2=75.6$m^2$& 0& 0& 25 & 500& 892k &261\\
2&5(BLE)&detection&8.8x8.6x3.2=75.6$m^2$& 0& 6& 100& 600&494k  &344\\
3&5(BLE)&counting &8.8x8.6x3.2=75.6$m^2$& 0& 6& 100& 600&1483k &181\\
\hline
4&4(BLE)/5(AP) &detection/counting&6.6x4x2.75=26.4$m^2$& 0& 2& 200 & 330&1500k&1000\\
5&4(BLE)/12(AP)&detection/counting &18x6x6=108$m^2$& 0& 5& 800 & 800&7000k&2000\\
\hline
\hline
\multicolumn{10}{c}{on average, each access point (AP) has 4 interfaces for the 2.4 GHz and 4 interfaces for the 5 GHz band.}
\end{tabular}
}
\caption{Dataset characteristics (\scriptsize{Datasets 1 to 3 use only RSSI measurements. Datasets 4 and 5 use both RSSI and CSI data}).}\label{tab:datasets}
\end{table*}

\section{Experimental results}\label{sec:res}
For evaluating the proposed methodology, an extensive set of experiments and analyses have been conducted. 
The goal is to evaluate which of the trained models leads to the best classification and prediction in the scenarios we consider.

\subsection{Characteristics of the analyzed datasets}
Table~\ref{tab:datasets} shows the characteristics of the datasets we collected in three different environments (i.e., a university classroom, a home living room, and an industrial laboratory) by adopting the setup described in Section~\ref{sec:arch}. 
Columns 1 to 4 contain the dataset ID, the number of radio signal transmitters distributed in the environment, i.e., BLE devices or WiFi access points (APs), the considered scenario, and the size of the environment. 
Columns 5 to 8 show the lowest and highest number of people inside the environment during the considered period of time, and the minimal and maximal distance between the transmitters and the receiver.
Finally, Columns 9 and 10 show the number of samples and the size of the dataset.

\subsubsection{Experimental environments}
We next describe the 3 environments we tested in our experiments.
In all of them, the positions of the emitters and receivers were fixed. 
The users participating in the experiment did not know the positions of the devices (emitters and receivers) and could not touch or modify them. 
Furthermore, they did not carry any devices that could interact with our setup, e.g., BLE transceivers.\\
 
\par\noindent\textbf{Datasets 1, 2 and 3:} The experiments corresponding to these datasets were carried out in a university classroom (8.8 m x 8.6 m x 3.2 m) with 15 working stations. 
They involved 6 subjects: 1 female (29 years, 1.58 m height) and 5 males (25-29 years, 1.75-1.95 m height). 
We used 1 smartphone as a receiver and 5 or 6 BLE beacons as senders, as specified in Table~\ref{tab:datasets}.\\

\par\noindent\textbf{Datasets 4 and 5:}
We collected the considered datasets, both RSSI and CSI. 
To the best of our knowledge, no dataset that provides both RSSI (from BLE communication) and CSI (from WiFi communication) data concerning occupancy detection and counting scenarios is publicly available.
These datasets aim to compare RSSI data of a BLE piconet to CSI data of an IEEE 802.11ac network (WiFi) for benchmarking purposes. 
In particular, to extract CSI data, we used a Raspberry Pi 4/B+ that interacts with the nearest APs of the environment. For the CSI data extraction, we used the Nexmon firmware patching framework~\cite{nexmon:project, gringoli:19}. 
We developed a custom software in C++ for recording and storing these data.
The experiments for Dataset 4 were carried out in a home living room (6.6 m x 4 m x 2.75m). They involved 2 subjects: 1 female (54 years, 1.66 m height) and 1 male (26 years, 1.80 m height). 
Instead, the tests for Dataset 5 were carried out in an \href{https://www.icelab.di.univr.it/}{Industrial Computer Engineering (ICE)} laboratory (18 m x 6 m x 6 m), containing one production line, several machines, and various devices, such as sensors and actuators, which communicate using different communication protocols.
Our tests for Dataset 5 involved 1 female  (27 years, 1.80m) and 4 males (25-29 years, 1.75-1.95 m height). 
We could control the packet rate and hence sampling frequency of the BLE data. On the other hand, we did not have control over the CSI sampling frequency, which was given by the (off-the-shelf) WiFi network. In particular, in datasets 4 and 5, the CSI data was perceived as 45 Hz, but highly depended on the connection utilization. Overall, all the datasets together contain 4 hours of collected data and require 4GB of storage.

\subsection{Occupancy detection}
In this section, we evaluate the proposed occupancy detection technique.
To adequately mimic as many of the different situations that occur during an everyday use of a room, we have carried out the following experiments. 
Subjects entered and left the environment in an undefined, random order, with the only constraint that each of them must stay in the environment for at least one minute.
Besides, they have carried out the following different activities.
i) All were standing still, 
ii) all were in motion simultaneously, 
iii) all were sitting simultaneously, 
iv) some were moving, while some were sitting, and 
v) in Dataset 5, subjects took a position in one of the working stations.


The achieved results are reported in Table~\ref{tab:mode_2}. 
The results were computed by processing Datasets 1, 2, and 3 as if they belonged to a single contiguous dataset.
The data contained in these 3 datasets are made up of 68\% non-empty environment instances (i.e., the room was occupied) and of 32\% empty environment instances (i.e., they represent an empty room).

We evaluated 5 different classification models implemented in the \href{https://github.com/B-HAR-HumanActivityRecognition/B-HAR_Baseline-Human-Activity-Recognition}{B-HAR} framework\cite{demrozi2021bhar} by using the RSSI features dataset, as depicted in Figure~\ref{fig:training}.
The outcome represents the environment status: \textit{false} when the environment is empty, \textit{true} if at least one person is in the environment. 
In Table~\ref{tab:mode_2}, Column 1 and 2 show the RSSI data sampling frequency and the used classification models, Columns 3 to 6 show the results in terms of specificity (S), recall (R), precision (P), and comprehensive accuracy (A).
Rows 2 to 6 presents the five different classification models trained and tested using the data collected at 200 Hz (i.e., the maximal sampling frequency of the acceleration sensor.).
\begin{table}[!ht]
\centering
\begin{tabular}{c| c|c|c|c||c}
\hline
\makecell{RSSI\\sampling}  &
Model                      & 
\makecell{Specificity\\(S)}&
\makecell{Recall\\(R)}     & 
\makecell{Precision\\(P)}  & 
\makecell{Accuracy\\(A)}\\
\hline
\hline
200 Hz&kNN  &98.53\% &99.07\% &99.07\% &99.07\%\\
200 Hz&WkNN &98.17\% &98.97\% &98.97\% &98.97\%\\
200 Hz&LDA  &99.83\% &99.70\% &99.70\% &99.70\%\\
200 Hz&QLDA &99.78\% &99.77\% &99.77\% &99.77\%\\
\rowcolor[gray]{0.8}
200 Hz&SVM  &99.91\% &99.92\% &99.91\% &99.92\%\\
\hline
\hline
 20 Hz&SVM & 86.42\% & 86.19\% & 86.11\% & 86.35\% \\
 45 Hz&SVM & 97.22\% & 97.07\% & 97.04\% & 97.18\% \\
100 Hz&LDA & 98.10\% & 98.03\% & 98.03\% & 98.04\% \\
\hline
\hline
\end{tabular}
\caption{Occupancy detection results (Dataset 1, 2 and 3).}\label{tab:mode_2}
\end{table}
Overall, the SVM model with a linear kernel achieved the most noticeable results, i.e., 99.92\% recall (empty environment), 99.91\% specificity (non-empty environment), 99.91\% precision, and 99.92\% accuracy.
Compared to all other models we considered, the SVM model requires higher computational capabilities; however, the Keras library~\cite{chollet2015keras} provides a Quasi-SVM model implementation for Android-based mobile devices, with sufficiently low computational complexity to be run on a smartphone.
Furthermore, to identify the best system configuration, the architecture was tested using different sampling frequencies.
Rows 7 to 9 show the most accurate (among all the five tested models) classification results corresponding to the 20 Hz, 45 Hz, and 100 Hz sampling frequencies, respectively. 
As shown, decreasing the frequency from 200 Hz to 45 Hz only slightly affect the accuracy of the model, while under 45 Hz the negative impact on RMSE and the MAE is more significant.

By examining the classification outcome in detail, we observed that the incorrectly classified samples are mostly related to the situation in which people inside the environment are all seated, regardless of their number.

It is worth mentioning that we have presented the results for \textit{real-time} detection, with a detection delay of around $\SI{1}{s}$. 
A longer allowed delay of, e.g., $\SI{30}{s}$, would be sufficient for most applications, such as controlling the ventilation and air conditioning. 
This would also allow for sampling windows of $\SI{30}{s}$, for which we expect a much higher detection accuracy, also when all subjects are sitting.

\subsection{Occupancy counting}
Occupancy counting is realized using dedicated regression algorithms, as already described. 
As for occupancy detection, we study the results for this scenario using Datasets 1, 2, and 3 with seven different regression models.
Table~\ref{tab:mode_3} presents the results obtained using both the raw and feature datasets over RSSI data for a sensor sampling frequency of 200 Hz (Rows 2 to 8). Rows 9 to 11 show the most accurate (among all the seven tested models) estimations for sampling frequencies of 20 Hz, 45 Hz, and 100 Hz.
The outcome is an estimation of the number of persons within the environment. 

Our results suggest that the Random Forest regression model achieves the best results on the feature dataset for all considered sampling frequencies (i.e., 20 Hz, 45 Hz, 100 Hz, and 200 Hz). 
As for the case related to occupancy detection, decreasing the frequency from 200 Hz to 45 Hz only slightly affects the accuracy of the model, while the negative impact on RMSE and the MAE is more significant under 45 Hz. Given that the maximum connection interval in BLE is $\SI{7.5}{ms}$, which corresponds to a frequency of roughly 133 Hz, it is expected that sampling rates between 200 Hz and 133 Hz lead to a similar connection interval and hence to the same accuracy. 

In summary, given a set of features based on RSSI measurements, the proposed occupancy counting system can estimate the number of people in the environment with an RMSE of 0.4 and an MAE of 0.3. In other words, in almost all cases, the Random Forest estimator can correctly identify the number of the environment occupants, with an error of at most $\pm$ 1 person.
\begin{table}[!ht]
\centering
\begin{tabular}{c|c||c|c||c|c}
\hline
RSSI      &
Regression                     &
\multicolumn{2}{c||}{Raw data} & \multicolumn{2}{c}{Features}   \\
sampling &model & RMSE & MAE & RMSE & MAE  \\
\hline
\hline
200 Hz&Gradient boosting   &0.9 &0.6 &0.4 &0.3  \\
\rowcolor[gray]{0.8}
200 Hz&Random forest       &0.7 &0.4 &0.4 &0.3  \\
200 Hz&Linear              &1.4 &1.0 &1.3 &1.0  \\
200 Hz&Ridge               &1.4 &1.0 &2.2 &4.1  \\
200 Hz&RANSAC              &1.8 &1.3 &3.1 &3.1  \\
200 Hz&Bayesian            &1.4 &1.0 &1.9 &1.9  \\
200 Hz&TheilSen            &1.9 &1.2 &2.0 &1.8  \\
\hline
\hline
 20 Hz&Random forest       &3.4 &3.1 &2.8 &2.2  \\
 45 Hz&Random forest       &1.9 &0.9 &1.4 &0.7  \\
100 Hz&Random forest       &1.1 &0.7 &0.8 &0.5  \\
\hline
\hline
\end{tabular}
\caption{Occupancy counting results (Datasets 1, 2 and 3).}\label{tab:mode_3}
\end{table}

When using only the raw dataset, we achieved an RMSE of 0.7 and an MAE of 0.4. Hence, the computation of the features is justified by the increased regression quality.
As for the occupancy detection scenarios, the estimation error is amplified when all people inside the environment are sitting.
It is worth noting that unlike many other existing works, in which a limited number of environmental situations are studied, our goal is to account for the most realistic environmental situations, into which a group of people perform different activities.

\begin{table*}[!ht]
\centering
\resizebox{\textwidth}{!}{
\begin{tabular}{c|c|cccc||cccc|||cccc||cccc}
\hline
\makecell{RSSI\\sampling} &
Model 
& S & R & P & A
& S & R & P & A
& S & R & P & A
& S & R & P & A\\
\hline
\hline
200 Hz&
kNN  &100\% &100\% &100\% &100\%
     &100\% & 98\% & 99\% & 98\%
     & 90\% & 91\% & 91\% & 91\%
     & 98\% & 99\% & 98\% & 98\%\\
200 Hz&
WkNN &100\% &100\% &100\% &100\%
     & 97\% & 98\% & 98\% & 98\%
     & 90\% & 91\% & 91\% & 91\%
     & 97\% & 99\% & 98\% & 99\%\\
200 Hz&
LDA  & 99\% & 99\% & 99\% & 99\%
     &100\% &100\% &100\% &100\%
     & 89\% & 89\% & 89\% & 89\%
     & 99\% & 99\% & 99\% & 99\%\\
200 Hz&
QLDA & 98\% & 99\% & 99\% & 99\%
     & 99\% & 98\% & 98\% & 98\%
     & 53\% & 58\% & 57\% & 57\%
     & 96\% & 96\% & 96\% & 96\%\\
\rowcolor[gray]{0.8}
200 Hz&
SVM  & 99\% & 99\% & 99\% & 99\%
     & 98\% & 99\% & 99\% & 99\%
     & 94\% & 94\% & 94\% & 94\%
     & 99\% & 99\% & 99\% & 99\%\\
\hline
\hline
 20 Hz&
SVM  & 87\% & 86\% & 87\% & 86\%
     & \multicolumn{4}{c|||}{CSI sampling }
     & 84\% & 84\% & 84\% & 84\%
     &\multicolumn{4}{c}{CSI sampling }\\
 45 Hz&
SVM  & 97\% & 96\% & 97\% & 96\%
     & \multicolumn{4}{c|||}{frequency is }
     & 92\% & 92\% & 92\% & 92\%
     & \multicolumn{4}{c}{frequency is }\\
100 Hz&
SVM  & 98\% & 98\% & 98\% & 98\%
     & \multicolumn{4}{c|||}{not settable.}
     & 93\% & 93\% & 93\% & 93\%
     & \multicolumn{4}{c}{not settable.}\\
\hline
\hline

\multicolumn{2}{c}{ }& 
\multicolumn{4}{c}{Dataset 4 RSSI-based} & \multicolumn{4}{c}{Dataset 4 CSI-based}&
\multicolumn{4}{c}{Dataset 5 RSSI-based} & \multicolumn{4}{c}{Dataset 5 CSI-based}
\end{tabular}}
\caption{Occupancy detection results (Datasets 4 and 5) - BLE vs. CSI.}\label{tab:mode_2_rssi_csi}
\end{table*}

\begin{table*}[!ht]
\centering
\resizebox{\textwidth}{!}{
\begin{tabular}{c|c|cc|cc||cc|cc|||cc|cc||cc|cc}
\hline
RSSI      &
Regression&
\multicolumn{2}{c|}{Raw Data} & \multicolumn{2}{c||}{Features}& \multicolumn{2}{c|}{Raw Data} & \multicolumn{2}{c|||}{Features}&
\multicolumn{2}{c|}{Raw data} & \multicolumn{2}{c||}{Features}& \multicolumn{2}{c|}{Raw data} & \multicolumn{2}{c}{Features}\\
sampling   & 
Model      & 
RMSE & MAE & RMSE & MAE  &  RMSE & MAE & RMSE & MAE & 
RMSE & MAE & RMSE & MAE  &  RMSE & MAE & RMSE & MAE \\
\hline
\hline
200 Hz &
Gradient boosting   &0.19&0.09&0.13&0.05
                    &0.13&0.05&0.11&0.04
                    &1.22&0.92&1.02&0.62
                    &\hlgray{0.31}&\hlgray{0.21}&\hlgray{0.28}&\hlgray{0.16}\\
200 Hz &
Random forest       &\hlgray{0.18}&\hlgray{0.05}&\hlgray{0.16}&\hlgray{0.04}
                    &\hlgray{0.08}&\hlgray{0.01}&\hlgray{0.04}&\hlgray{0.01}
                    &\hlgray{0.91}&\hlgray{0.47}&\hlgray{0.91}&\hlgray{0.52}
                    &0.81&0.34&0.31&0.25\\ 
200 Hz &
Linear              &0.46&0.36&0.33&0.22
                    &0.81&0.48&0.92&0.29
                    &1.62&1.34&1.31&1.21
                    &0.91&0.61&1.03&0.52\\
200 Hz &
Ridge               &0.46&0.36&0.34&0.22
                    &0.57&0.32&0.61&0.38
                    &1.68&1.37&1.71&1.34
                    &1.41&0.97&1.26&0.66\\
200 Hz &
RANSAC              &0.45&0.37&0.35&0.25
                    &0.58&0.34&0.60&0.40
                    &1.80&1.45&1.75&1.38
                    &1.43&0.94&1.34&0.72\\
200 Hz &
Bayesian            &0.46&0.37&3.40&2.30
                    &0.57&0.34&9.12&4.22
                    &1.65&1.34&7.10&4.72
                    &1.71&1.20&1.58&1.06\\
200 Hz &
TheilSen            &47.0&1.37&14.7&29.9
                    &36.1&3.62&1.11&0.31
                    &1.85&1.47&6.55&7.21
                    &15.4&10.5&20.4&14.1\\
\hline
\hline
 20 Hz &
Random forest       &3.78&2.52&3.48&2.24
                    &\multicolumn{4}{c|||}{CSI sampling }
                    &7.88&6.17&7.52&6.10
                    &\multicolumn{4}{c}{CSI sampling }\\
 45 Hz &
Random forest       &0.34&0.15&0.30&0.15
                    &\multicolumn{4}{c|||}{frequency is }
                    &2.27&1.99&2.30&2.11
                    &\multicolumn{4}{c}{frequency is }\\
100 Hz &
Random forest       &0.24&0.12&0.21&0.11
                    &\multicolumn{4}{c|||}{not settable.}
                    &1.95&1.73&1.92&1.58
                    &\multicolumn{4}{c}{not settable.}\\
\hline
\hline
\multicolumn{1}{c}{ }& 
\multicolumn{4}{c}{Dataset 4 RSSI-based} & \multicolumn{4}{c}{Dataset 4 CSI-based}& 
\multicolumn{4}{c}{Dataset 5 RSSI-based} & \multicolumn{4}{c}{Dataset 5 CSI-based}
\end{tabular}}
\caption{Occupancy counting results (Datasets 4 and 5) - BLE vs. CSI.}\label{tab:mode_3_rssi_csi}
\end{table*}

\subsection{RSSI vs. CSI}
As already described, the results presented so far concern BLE networks computed using RSSI data.
In this section, we compare the performance when using BLE against the performance when using WiFi/CSI data.
For this purpose, we created Datasets 4 and 5, which contain both RSSI and CSI data collected in the same environments.

Table~\ref{tab:mode_2_rssi_csi} shows the results we obtained for Datasets 4 and 5 in the occupancy detection scenario. 
Column 2 lists the five considered classifiers. 
Columns 3 to 6 (RSSI-based) and 7-10 (CSI-based) show the results when using Dataset~4, while Columns 11 to 14 (RSSI-based) and 15-18 (CSI-based) show the results for Dataset 5.
Similarly, Table~\ref{tab:mode_3_rssi_csi} depicts the results obtained for Datasets 4 and 5 for occupancy counting.  
In particular, the upper parts of Tables \ref{tab:mode_2_rssi_csi} and \ref{tab:mode_3_rssi_csi} present the achieved results obtained using data sampled at 200 Hz (w.r.t. acceleration data over BLE) and 45 Hz (w.r.t. CSI data).
Similarly, the bottom parts present the achieved results for data sampled at 20 Hz, 45 Hz, and 100 Hz.
We can summarize the achieved results by comparing CSI vs. RSSI as follows:
\begin{itemize}
    \item As expected for both RSSI and CSI, the detection/prediction for Dataset 4 works more accurately than for Dataset 5, mainly due to the environment size and shape. Moreover, various electrical machines were present and potentially interfering in the environment of Dataset 5. Besides, the larger height of the environment has negatively affected the results;
    \item Running the classifiers on the features dataset achieved more reliable results than when using the raw dataset;
    \item In our experiments, on average, the SVM classifier achieved the most reliable results concerning the occupancy detection problem. The Random Forest regression algorithm achieved the best results concerning the occupancy counting problem;
    \item The more complex and comprehensive CSI data leads to an only marginally higher detection accuracy than our proposed BLE-based system, as shown in Tables~\ref{tab:mode_2_rssi_csi} and~\ref{tab:mode_3_rssi_csi}. Hence, there is little benefit in setting up a more complex and expensive WiFi-based detection system. It is worth mentioning that in the studied environments, the number of information sources for CSI was larger than for BLE (i.e., four vs. five in Dataset 4 and four vs. twelve in Dataset 5).
\end{itemize}
The overall accuracy of the proposed methodology, among all the three tested environments, when taking into account only the most accurate classifiers, is 97.97\% for occupancy detection. The RMSE for occupancy counting on the average of all the five tested datasets is 0.32.

\subsection{Sampling frequency and energy consumption}
The sensor sampling frequency used for the results presented above were $\SI{20}{Hz}$, $\SI{45}{Hz}$, $\SI{100}{Hz}$, and $\SI{200}{Hz}$ for BLE. For WiFi, we recorded packets with an average frequency of $\SI{45}{Hz}$.
Concerning BLE devices, a lower sampling frequency would also reduce the energy consumption of the BLE transmitters and - to a minor extent - also of the receiver. 
Since battery life also affects the usability of our approach, we study its energy consumption in detail in this section. 
We first discuss the energy characteristics of the hardware we have used.
We evaluate the average RAM and memory demand, as well as the energy consumption of the transmitters (i.e., Nordic Thingy 52 with a 64 MHz Cortex M4 MCU, 512 Kb Flash, 64 Kb RAM, and a Battery of 1440 mAh) and the receiver (Oneplus 6 with a Qualcomm SDM845 Snapdragon 845, featuring an Octa-core 4x2.8 GHz Kryo 385 Gold \& 4x1.7 GHz Kryo 385 Silver CPU, 128 GB memory, 8 GB RAM, with Android 10, OxygenOS 10.3.7, and a 3300 mAh Battery).\\

\par\noindent\textbf{Energy consumption of the transmitters: }
When operating using a sensor sampling frequency of 200 Hz, the BLE devices can efficiently operate for more than five days without recharging their battery. 
A lower sampling frequency would further extend their battery lifetime significantly~\cite{kindt:20a}. 
Therefore, a sampling frequency of $\SI{45}{Hz}$ represents a good trade-off between the accuracy of detection/classification and battery runtime.\\

\par\noindent\textbf{Energy consumption of the receiver: }
The energy consumption of the smartphone is only marginally affected by the energy needed for BLE communication~\cite{kindt:20b}. 
For our smartphone, when sampling at 200 Hz, we found the following average values per hour: 6\% of CPU usage,  154 mA concerning battery consumption, 122 Mb of memory, 100 Mb of RAM, and 0.34\% of data loss. 
Instead, sampling at 45 Hz, we found the following average values per hour: 1\% of CPU usage, 94 mA concerning battery consumption, 74 Mb of memory, 90 Mb of RAM, and 0.12\% of data loss.

\section{Concluding remarks and future work}\label{sec:conc}
Occupancy detection and occupancy counting provide important information for smart cities and smart building environments in several scenarios. 
However, existing solutions have many limitations, mainly related to high economic cost, low accessibility, high computational requirements, difficulties of installation, and lack of ubiquitousness. 
This paper presented a pattern recognition-based methodology that uses low-cost BLE communication technology for occupancy detection and counting.
It can be retrofitted into any environment with negligible installation effort. 
Different regression and classification algorithms were used, achieving promising results in different environments. 
In particular, occupancy can be detected, taking into account only the best classifier, with an average accuracy of 97.97\% over all datasets. 
The number of people in a room can be estimated with an average RSME/MAE of 0.32/0.28 people.
We showed that our methodology working on BLE RSSI data achieves practically the same accuracy as WiFi/CSI-based approaches do. 
At the same time, it comes with a much lower cost and installation effort.

The datasets we used in our experiments have been created with a relatively limited number of persons. This is due to regulations in response to the ongoing SARS-CoV-2 pandemic, which prevents us from placing more persons into the same environment.
While detecting the occupancy is expected to work even more reliably when a larger number of persons is present in a room, the occupancy counting method needs to be evaluated further when the SARS-CoV-2 restrictions have been eased.

Our objective for future research is to reduce the number of senders while maintaining the same performance. Simultaneously, it is desirable to reduce the dependence of the methodology on the sender positions and the explicit knowledge of their distances from each other.
This might be done, e.g., by estimating their distance automatically.
Moreover, we aim to use our proposed system and the created classification and regression models as substantial models to move forward the transfer learning research area to recognize an environment's context without performing a preliminary offline data collecting phase.
Some smartphone models, such as the Honor 7 or Samsung Galaxy S5 and S7, fail to keep up with the required sampling rates. The limitations of such devices are mainly related to the operating system's version. For limiting the power consumption, some custom OS and HW versions do not allow an application to extract RSSI values at frequencies higher than 40 Hz. On the contrary, devices that do not present such limitations are, for example, all Oneplus devices and the Samsung Galaxy S9. In future research, we attempt to address this by further reducing the sampling rate requirements, making it compatible with literally all smartphone models.

Finally, our current system carries out the online classification on a server. However, our detection algorithms are light-weight enough to be run on a smartphone. Hence, the server is only needed for the learning phases, while the online detection can be done on the smartphone in the future. 

\bibliographystyle{IEEEtran}
\bibliography{IEEEabrv,biblio}{}

\vspace{-1.2cm}
\begin{IEEEbiography}
[{\includegraphics[width=0.9in,height=1.25in,clip,keepaspectratio]{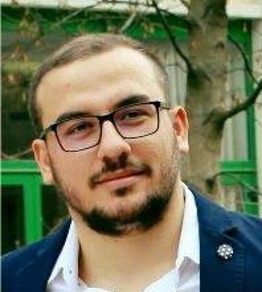}}]{\textbf{Florenc Demrozi,}} Ph.D in computer science, IEEE member, received 
the Ph.D. degree in Computer Science in 2020. He is currently a Postdoctoral researcher and Temporary Professor at the Department of Computer Science, University of Verona, Italy, working on Human Activity Recognition (HAR), Ambient Intelligence (AmI), Ambient Assisted Living (AAL) and Internet of Things (IoT).
\end{IEEEbiography}
\vspace{-1.2cm}
\begin{IEEEbiography}
[{\includegraphics[width=1in,height=1.1in,clip]{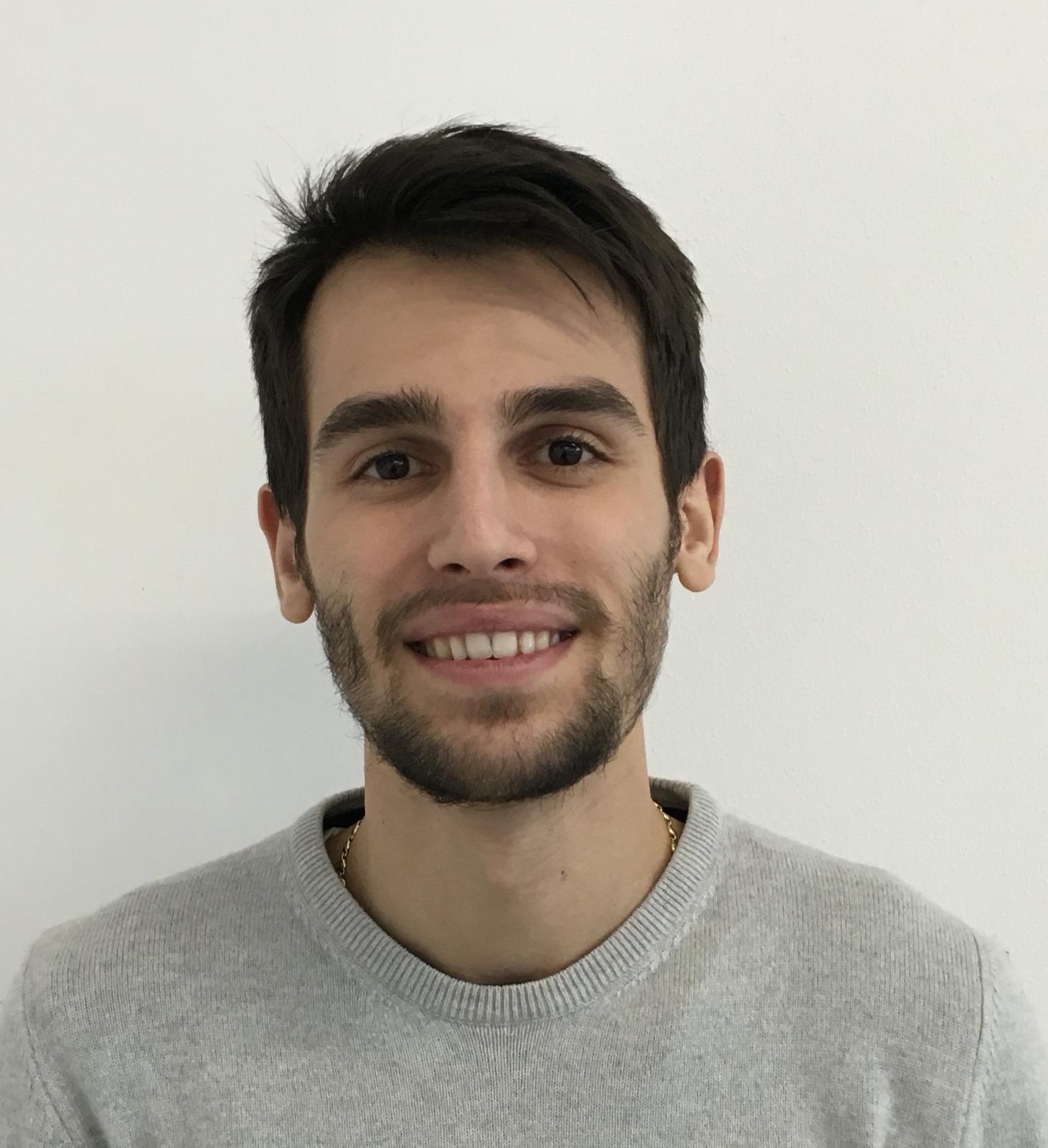}}]{\textbf{Cristian Turetta,}} received the B.S. and M.E. degrees in Computer Science and Engineering from the University of Verona, Italy, respectively in 2017 and 2020. He is currently a research fellow at the Department of Computer Science, University of Verona, Italy working on Ambient Assisted Living (AAL), Internet of Things (IoT) and IoT Security.
\end{IEEEbiography}
\vspace{-1.2cm}
\begin{IEEEbiography}
[{\includegraphics[width=1in,height=1in,clip]{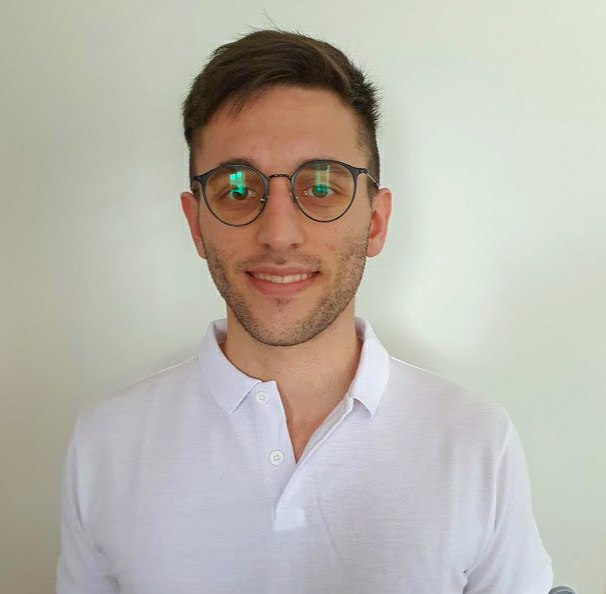}}]{\textbf{Fabio Chiarani,}} received the B.S. degrees in Computer Science from the University of Verona, Italy, in 2020. He is currently a M.E. student in Computer Science and Engineering from the University of Verona, Italy.  His main interests focus on Edge Computing and IoT.
\end{IEEEbiography}
\vspace{-1.2cm}
\begin{IEEEbiography}
[{\includegraphics[width=0.9in,height=1.15in,clip,keepaspectratio]{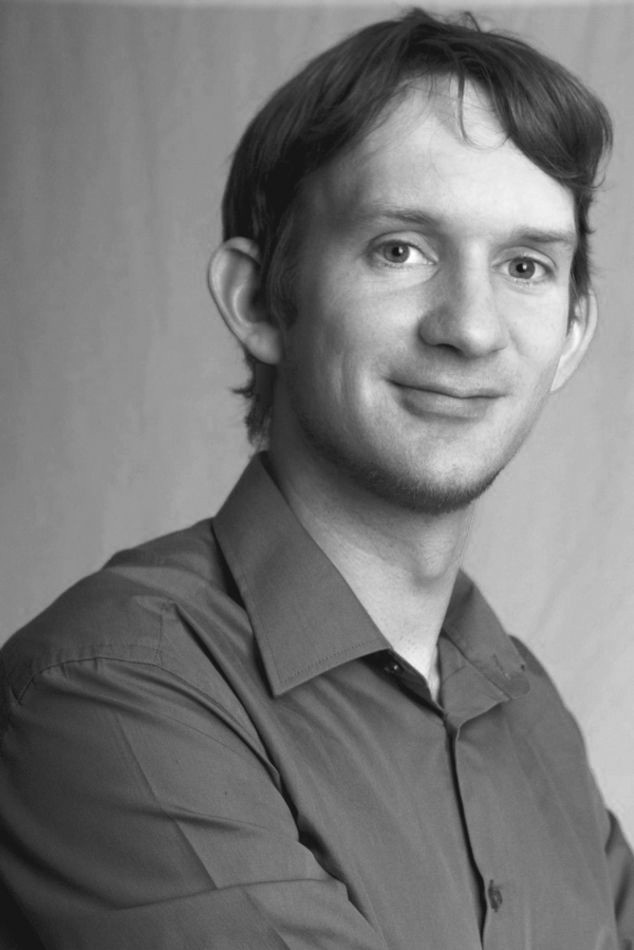}}]{\textbf{Philipp H. Kindt,}} received his Ph.D. in Electrical Engineering from the Technical University of Munich (TUM) in 2019. He is currently an assistant professor (``Juniorprofessor'') of pervasive computing systems at the Chemnitz University of Technology. His research interests are wireless communication, mobile computing and the IoT.
\end{IEEEbiography}
\vspace{-1.2cm}
\begin{IEEEbiography}
[{\includegraphics[width=0.9in,height=1.25in,clip,keepaspectratio]{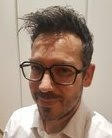}}]{\textbf{Graziano Pravadelli,}} PhD in computer science, IEEE senior member, IFIP 10.5 WG chair, is full professor of information processing systems at the Computer Science Department of the University of Verona (Italy) since 2018. In 2007 he cofounded EDALab s.r.l., an SME working on the design of IoT-based monitoring systems. His main interests focus on system-level modeling, simulation and semi-formal verification of embedded systems, as well as on their application to develop IoT-based virtual coaching platforms for people with special needs. In the previous contexts, he collaborated in several national and European projects and he published more than 120 papers in international conferences and journals.
\end{IEEEbiography}
\end{document}